\documentstyle[12pt]{article}
\newcommand{\be}{\begin{equation}}
\newcommand{\ee}{\end{equation}}

\newcommand{\bea}{\begin{eqnarray}}
\newcommand{\eea}{\end{eqnarray}}

\newcommand{\bean}{\begin{eqnarray*}}
\newcommand{\eean}{\end{eqnarray*}}

\newcommand{\ka}{\kappa_{ \gamma}}
\newcommand{\la}{\lambda_{ \gamma }}

\newcommand{\ba}{\begin{array}}
\newcommand{\ea}{\end{array}}
\newcommand{\lsim}{{\;\raise0.3ex
\hbox{$<$\kern-0.75em\raise-1.1ex\hbox{$\sim$}}
\;}}
\newcommand{\gsim}{{\;\raise0.3ex
\hbox{$>$\kern-0.75em\raise-1.1ex\hbox{$\sim$}}
\;}}

\newcommand{\DE}{\Delta}
\newcommand{\g}{\gamma}

\newcommand{\th}{\theta}

\newcommand{\ie}{{\it i.e. }}
\newcommand{\eg}{{\it e.g. }}

\newcommand{\lr}{{left-right symmetric model}}

\begin{document}

\begin{titlepage}

\mbox{}\vspace*{-1cm}\hspace*{8cm}\makebox[7cm][r]{\large  HU-SEFT R 1994-16}
\mbox{}\vspace*{-0cm}\hspace*{9cm}\makebox[7cm][r]{\phantom  {HU-SEFT R
1993-17}}
\vspace*{0.5cm}
 \hspace*{10.9cm} \makebox[4cm]{(hep-ph/yymmxxx)}
\vfill

\Large

\begin{center}
{\bf  Tests of gauge boson couplings
 in polarized  $e^-\g$ collisions}

\bigskip
\normalsize
{{\rm Martti
Raidal}}\\[15pt]{\it Research Institute for High Energy Physics, \\
P.O.Box 9, FIN-00014, \\
University of Helsinki}

{November 1994}

\bigskip

\vfill

\normalsize

{\bf\normalsize \bf Abstract}
\end{center}

\normalsize

Single $W$-boson  production  in $e^-\g$ collisions with
polarized beams is investigated. Helicity amplitudes for general
couplings are derived and their properties are discussed.
The results are applied to the Standard
Model (SM) and the left-right model.
In the framework of  SM the
updated estimates of the measurement precision of
 photon anomalous coupling parameters $ \ka,$ $ \la$ at the NLC with
$ \sqrt{s_{ ee }}=500$GeV are obtained.
The production of right-handed gauge bosons $ W^-_{ 2 }$ in these
collisions is also analysed.

\end{titlepage}
\normalsize

\section{Introduction}

In addition to the  electron-positron option,
the electron-electron and electron-photon collision modes
of the Next Linear Collider (NLC) are also
technically realizable  \cite{orava}.
During the recent years the physics potential of
the  latter options has
been under intense study.  While $e^-e^-$ collisions
have been found to be particularly suitable for
the study of possible  lepton number violating
 phenomena \cite{cuy},  the $e^-\gamma$ operation
mode will also be well motivated
from the  point of view of new physics.

So far, the $ e^-\gamma$ collisions have been studied using the photon spectrum
of  classical Bremsstrahlung.
In the linear collider it will be possible to obtain high luminosity
photon beams by  backscattering intensive laser pulses off the electron beam
\cite{ginz} without considerable losses in the  beam energy and with
very high polarizability and monochromaticity \cite{telnov}.
 This possibility makes the  $ e^-\gamma$ collisions well
suited for testing the  Higgs sector of
the Standard Model (SM) \cite{higgs}, as well as for studying
supersymmetric theories \cite{susy}.
Moreover, the $ e^-\gamma$ collision mode would be
 ideal for studying heavy gauge boson
production processes \cite{eg,choi,king},
since the initial state photon provides us with a possibility
 to probe directly  the gauge boson self-interactions.

In this paper we study a single massive vector boson production in
$ e^-\gamma $ collision,

\be
e^-\gamma\rightarrow W^-N,
\label{prot}
\ee
for any combination of beam polarization.
Here $ W^-$ may stand for the ordinary SM charged vector boson
$ W^-_1$ and $ N$ for the massless Dirac
electron neutrino $ \nu_e$. However, we do not restrict ourselves  only
to this case, since a wide class of models beyond
the SM predicts a existence of
new heavy vector bosons and massive neutrinos. For example,
in the left-right symmetric theory \cite{lr} based on the
gauge group $ SU(2)_R\times SU(2)_L\times U(1)_{ B-L }$
the vector boson may  also be a heavy  right-handed
weak boson $ W_2$.  There are two Majorana neutrinos in this model,
which, in principle, can be mixed to give one heavy and one light mass
eigenstate.
The present lower limit for the mass of $ W_2$ coming from TEVATRON is
$M_{W_2}\geq 652$ GeV \cite{mass}, so that the right-handed boson production
will be kinematically forbidden in the initial phase of   NLC, where the
$ e^-e^+$ center of mass energy is planned to be $ \sqrt{s_{ ee }}=500$ GeV.
At the final phase
of  NLC, however, reaction (\ref{prot}) could be kinematically allowed and
favoured compared with, \eg, the  $ W_2$ pair production in $ e^-e^+$
collisions, since the mass of
 heavy Majorana neutrino  could be smaller
than the mass of  $ W_2.$ In the case of a sizeable
mixing  between the light, predominantly left-handed and the heavy,
predominantly right-handed Majorana neutrinos the study of
  process (1)  may extend the kinematical
discovery range of $ W_2 $ almost up to the energy $ \sqrt{s_{ e\gamma} }.$

 There are two Feynman diagrams
 contributing at the tree level to  reaction (\ref{prot}) (see Fig. 1).
One of them, the u-channel diagram, involves a triple gauge boson coupling
making the  process  suitable for testing the non-Abelian gauge
structure of the theory. A particularly interesting feature of  process
(1) is that it is sensitive {\em only} to the possible
anomalous coupling of the  photon,
allowing one  to discriminate between the photon anomalous coupling
and the  anomalous coupling of  massive
neutral gauge boson $ Z^0.$
In a gauge theory the total cross section of  reaction (\ref{prot})
approaches a constant value at  high energies. Any deviation from the
gauge form of the triple boson coupling would spoil the good high
energy behaviour  and lead to
 a violation of unitarity at some energy.
At the energies of  NLC these deviations are expected to be small,
which will make their detection difficult.
However, using polarized initial state particles one can
enhance these effects.

In what follows  we shall investigate  reaction (\ref{prot}) for
polarized beams  taking into account the final state polarization
measurements. We shall derive the helicity amplitudes assuming a general form
of the relevant couplings and discuss their  behaviour, particularly at
high energies and with anomalous couplings.
We shall apply the general amplitudes in the case of two sample
theories: the SM and the left-right model. We shall give an  updated
estimate  for the  precision of the anomalous triple boson
coupling measurement at the first stage of  NLC
by analysing five different observables.
We shall also study the  discovery
potential of the right-handed vector boson in the final phase
of  NLC via reaction (1).

The paper is organized as follows. In Section 2 we derive the helicity
amplitudes and discuss the properties of the cross sections.
In Section 3 we analyse the sensitivity of  NLC to the
triple boson coupling in the framework of the SM.
In Section 4 we study  reaction (1) in the case of the left-right model.
A summary is given in  Section 5.

\section{Helicity amplitudes}

In a gauge theory the transverse and longitudinal  components of
gauge bosons have  different origins related to the
 different aspects  of theory. The
longitudinal components of  massive vector bosons
exist  due to the Higgs mechanism,
while the existence of the transverse components is dictated
by the gauge invariance. Therefore, in order to obtain more information
about the properties of  gauge boson self-interactions, it would  be
 useful  to
investigate reaction (1) by taking into account the polarization of
particles. For this purpose we  derive the  helicity amplitudes
of  the process.

We do not  restrict ourselves to any particular model but
define the relevant couplings as general as possible.
We assume that the charged current interaction is of the  form
\be
{\cal L}^{cc}=\frac{g}{2\sqrt{2}}\overline{N}\gamma_{ \mu }
(A + B\gamma_5)eW
^{ +\mu } + h.c.,
\ee
where $ g$ is a coupling constant and $ A$ and $ B$ are parameters which
allow us to choose an
arbitrary vector-axial-vector structure for the interaction.
The parameters $ A $ and $ B$
enable us also to incorporate possible
mixings  of the particles.

The most general $ CP$-conserving
$ \gamma WW$ interaction allowed by the electromagnetic gauge invariance
is of the form \cite{hagi}
\be
{\cal L}_{ \gamma WW }=-ie(W^{\dagger}_{ \mu\nu }W^{\mu}A^{\nu} -
W^{\dagger}_{ \mu }A_{\nu}W^{\mu\nu} + \kappa_{ \gamma }
W^{\dagger}_{ \mu }W_{\nu}F^{\mu\nu} + \frac{\lambda_{\gamma}}{M^2_W}
W^{\dagger}_{ \tau\mu }W^{\mu}_{\nu}F^{\nu\tau}),
\label{triple}
\ee
where $ W_{ \mu\nu }=(\partial_{ \mu } - ieA_{ \mu })W_{ \nu } -
(\partial_{ \nu } - ieA_{ \nu })W_{ \mu }$ and
$ F_{ \mu\nu }=\partial_{ \mu }A_{ \nu } - \partial_{ \nu }A_{ \mu }.$
The coefficients $ \kappa_{ \gamma }$ and $ \lambda_{ \gamma }$ are
related to the magnetic moment $ \mu_W$ and the electric quadrupole
moment $ {\cal Q}_W$ of  $ W$ according to
\[\mu_W=\frac{e}{2M_W}(1 + \kappa_{ \gamma } + \lambda_{ \gamma }),\]
\[{\cal Q}_W=-\frac{e}{M^2_W}(\kappa_{ \gamma } -  \lambda_{ \gamma }).\]
In a gauge theory at  tree level the coefficients have the values
$ \ka=1$ and $ \la=0$.

In the case of  the s-channel diagram we also need the $ \gamma e^-e^-$
vertex, which is assumed to
have the same form as in QED.

In our calculation we have neglected the electron mass.  Accordingly,
the longitudinally polarized electrons coincide with their left- and right-
handed chirality states denoted by $ \lambda=-\frac{1}{2}$ and
$ \lambda=\frac{1}{2},$ respectively. The electron momentum $ q$ is taken to
be along the z-axis, $ q^{\mu}=(|q|,0,0,|q|),$
and the momentum of the final state $ W^-$
 is given by $ k^{\mu}=(E_W,|k|\sin\theta,0,|k|\cos\theta).$
We describe the polarized initial state photon by the polarization vectors
\be
\epsilon^{\mu}_{photon }(k_1,\tau=\pm 1) = \frac{1}{\sqrt{2}}(0, \tau, -i, 0),
\ee
and the polarization states of the
final state $ W^-$ by the polarization vectors
\[
 \epsilon^{\mu *}_{W }(k,\tau=\pm 1)  =
\frac{1}{\sqrt{2}}(0, -\tau\cos\theta,i, \tau\sin\theta),
\]
\be
\epsilon^{\mu *}_{W }(k,\tau=0)  = \frac{1}{\sqrt{M_W}}(|k|, E_W\sin\theta,
 0, E_W\cos\theta).
\ee
The spinor $ \overline{u}_N,$ which
describes a neutrino with mass $ M_N$ and momentum
$ p^{\mu} = (E_N, \vec{p}),$ $\vec{p}=-\vec{k},$
is quantized with respect to the positive  z-axis.

Now we write down the helicity amplitudes of  process (1).
Using the notation given above the helicity amplitudes can be
written as
\be
F_{ \lambda\lambda'}^{\tau\tau'}=i\frac{eg(A+2\lambda B)}{2\sqrt{2}}
\overline{u}_N(p,\lambda')T_{ \mu\nu }u_e(q,\lambda)\epsilon_{ photon
 }^{\nu}(k_1,\tau)\epsilon_W^{\mu *}(k,\tau'),
\ee
where the tensor $ T_{ \mu\nu }$ is the sum of the two terms corresponding to
the s- and u-channel diagrams. The first lower and the first upper indices in
$ F_{ \lambda\lambda'}^{\tau\tau'}$ describe the polarization state of
the electron and photon and the second pair describes
the polarization states of the
final state  neutrino and $ W$-boson, respectively.
Tensor  $ T_{ \mu\nu }$
is of the form
\be
T_{ \mu\nu }  =  \frac{\gamma_{ \mu }(q\!\!\!/ +
k_1\!\!\!\!\!/\/ )\gamma_{ \nu }}{s}
 -\frac{1}{u-M^2_W}(\gamma^{\rho}-\frac{(p\!\!/-q\!\!\!/)(k_1-k)
^{\rho}}{M^2_W})\Gamma_{ \nu\mu\rho },
\ee
where
\bea
 \Gamma_{ \nu\mu\rho } & = & -2k_{ \nu }g_{ \mu\rho }-(1+\ka-\la)k_{1\mu  }
g_{ \nu\rho } + (k+(\ka-\la)k_1)_{ \rho }g_{ \mu\nu } \nonumber \\
& & +\frac{\la}{M^2_W}(k+k_1)_{ \rho }((k\cdot k_1)g_{ \mu\nu } - k_{ \nu }
k_{ 1\mu }),
\eea
with $ -ie\Gamma_{ \nu\mu\rho }$ being  the $ \gamma WW$
vertex.

The amplitudes read as ($ \lambda=\pm\frac{1}{2}$ and
$\tau,\tau'=\pm1,$ longitudinal $ W$'s are denoted by 0):
\bea
F_{ \lambda\lambda }^{\tau\tau'} & = & -i\frac{eg}{8}(A+2\lambda B)
\sqrt{\sqrt{s}(E_N+M_N)}
\left((1+\frac{|p|}{E_N+M_N})\frac{(1+2\lambda \tau)
(1+\tau\tau')}{\sqrt{s}}\sin\frac{\theta}{2}\right. \nonumber  \\
& & + \frac{\tau\tau'}{u-M^2_W}
 \left\{(1+\frac{|p|}{E_N+M_N})
\left[(G_2(1+2\lambda\tau')-G_1(1+2\lambda\tau))\cos\frac{\theta}{2} -
G_3\sin\frac{\theta}{2}\right]\right. \nonumber \\
& & \left.\left.-(1-\frac{|p|}{E_N+M_N})G_4
\sin\frac{\theta}{2}\right\}\right), \nonumber \\
 F_{ \lambda-\lambda }^{\tau\tau'} & = & -i\frac{eg}{8}(A+2\lambda B)
\sqrt{\sqrt{s}(E_N+M_N)}
\left((1-\frac{|p|}{E_N+M_N})\frac{(1+2\lambda \tau)
(\tau-\tau')}{\sqrt{s}}\cos\frac{\theta}{2}\right. \nonumber  \\
& & + \frac{2\lambda\tau\tau'}{u-M^2_W}
 \left\{(1-\frac{|p|}{E_N+M_N})
\left[(G_1(1+2\lambda\tau)+G_2(1-2\lambda\tau'))\sin\frac{\theta}{2} -
G_3\cos\frac{\theta}{2}\right]\right. \nonumber \\
& &  -\left.\left.  (1+\frac{|p|}{E_N+M_N})G_4
\cos\frac{\theta}{2}\right\}\right), \nonumber \\
 F_{ \lambda\lambda }^{\tau 0} & = & -i\frac{eg}{4\sqrt{2}M_W}(A+2\lambda B)
\tau\sqrt{\sqrt{s}(E_N+M_N)}
\left((1+\frac{|p|}{E_N+M_N})\frac{(1+2\lambda\tau)
(E_W+|p|)}{\sqrt{s}}\cos\frac{\theta}{2}\right. \nonumber  \\
& & - \frac{1}{u-M^2_W}
 \left\{(1+\frac{|p|}{E_N+M_N})
\left[G_1^0(1+2\lambda\tau)\cos\frac{\theta}{2} +
(G_2(E_W+|p|) +
G_3^0)\sin\frac{\theta}{2}\right]\right. \nonumber \\
& & +\left.\left.  (1-\frac{|p|}{E_N+M_N})G_4^0
\sin\frac{\theta}{2}\right\}\right), \nonumber \\
 F_{ \lambda-\lambda }^{\tau 0} & = & -i\frac{eg}{4\sqrt{2}M_W}(A+2\lambda B)
\sqrt{\sqrt{s}(E_N+M_N)}
\left((1-\frac{|p|}{E_N+M_N})\frac{(1+2\lambda\tau)
(E_W-|p|)}{\sqrt{s}}\sin\frac{\theta}{2}\right. \nonumber \\
& & + \frac{2\lambda\tau}{u-M^2_W}
 \left\{(1-\frac{|p|}{E_N+M_N})
\left[G_1^0(1+2\lambda\tau)\sin\frac{\theta}{2} +
(G_2(E_W-|p|) -
G_3^0)\cos\frac{\theta}{2}\right]\right. \nonumber \\
& & -\left.\left.  (1+\frac{|p|}{E_N+M_N})G_4^0
\cos\frac{\theta}{2}\right\}\right),
\label{hel}
\eea
where the factors $ G_i$ and $ G^0_i$   are defined as:
\bean
G_1 &=& -\frac{1+\ka-\la}{2}\sqrt{s}\sin\theta, \\
G_2 &=& 2|p|\sin\theta, \\
G_3 &=&
\sqrt{s}\left\{(1+\ka-\la)(\cos\theta-\tau\tau')+\frac{\la}{M^2_W}\left[
(M^2_W-u)(\cos\theta-\tau\tau')+\sqrt{s}|p|\sin^2\theta\right]\right\}, \\
G_4 &=& -\frac{1-\ka+\la}{2}\frac{M_N}{M_W^2}\left[
(M^2_W-u)(\cos\theta-\tau\tau')+\sqrt{s}|p|\sin^2\theta\right], \\
G^0_1 &=& -\frac{1+\ka-\la}{2}\sqrt{s}(|p|+E_W\cos\theta), \\
G^0_3 &= & \sqrt{s}\left\{-(1+\ka-\la)E_W\sin\theta +\frac{\la}{M^2_W}\left[
(u-M^2_W)E_W +\sqrt{s}|p|(|p|+E_W\cos\theta)\right] \sin\theta\right\}, \\
G^0_4 &=& -\frac{1-\ka+\la}{2}\frac{M_N}{M^2_W}\sin\theta\left[
(u-M^2_W)E_W +\sqrt{s}|p|(|p|+E_W\cos\theta)\right] .
\eean

All the amplitudes are proportional to the factor
$ A + 2 \lambda B$ which makes the amplitudes  vanish if the polarizations
do not match with  the chiral structure of the
charged current interaction. In the amplitudes
of  type $ F_{ \lambda-\lambda }^{\tau\tau'}$
the flip of the neutrino helicity
is controlled by the factor
$ (1-|p|/(E_N+M_N)),$ which makes these amplitudes  vanish in the case of
a massless neutrino.
Because of the annihilation into a massless fermion,
the s-channel contribution is nonzero only in the case when the
electron and  photon beams are similarly polarized.

In   SM one  has
pure (V-A) current (\ie $ A=-B=1$) and a massless Dirac neutrino,
which fixes the electron and neutrino polarization states to
$ \lambda=-1/2.$
If we assume a pure gauge model triple boson coupling then
the amplitudes $ F_{-\frac{1}{2}-\frac{1}{2} }^{-1 +1}$ and
 $ F_{-\frac{1}{2}-\frac{1}{2} }^{-1\;\;\: 0}$ vanish identically.

In the case of the left-right model
there is also a
 (V+A) current (fixing $ A=B=1$ and $ \lambda=1/2$) in addition to the
SM one and a massive Majorana
neutrino.  Again, without anomalous triple boson coupling, two of the
helicity amplitudes,  $ F_{+\frac{1}{2}+\frac{1}{2} }^{+1 -1}$ and
$ F_{+\frac{1}{2}-\frac{1}{2} }^{+1 +1},$ vanish.
Let us also note that
the terms in the helicity amplitudes  proportional to the neutrino
mass, \ie the terms with the coefficients $ G_4$ and $ G_4^0$, are
nonzero only if $ \ka$ and $ \la$  differ from their gauge model
values. This implies that the effects of the anomalous triple boson
coupling are  enhanced in the case of a massive neutrino
compared with the massless case.

For determining the cross sections from the helicity amplitudes
we have assumed 100\% longitudinally
 polarized  electron and linearly polarized photon beams. This is,
of course, an approximation, since
in practice the polarizations will never be ideal and one has to employ a
 density matrix giving the polarization parameters of the beams.

An interesting feature of  reaction (1) is that in a gauge
theory  the total cross section for polarized beams
is approaching a constant value,
\be
\sigma\rightarrow\frac{e^2g^2}{8\pi M_W^2},
\label{cr}
\ee
 at high energies. The only contributions which remain in this limit
are the ones corresponding to the case where
the photon and $ W$ are both polarized the same way. The other
terms  decrease very rapidly with energy.
  For example, in the SM   both  cross sections
$ \sigma_{-\frac{1}{2}-\frac{1}{2} }^{-1\; -1}$
and
$ \sigma_{-\frac{1}{2}-\frac{1}{2} }^{+1\; +1}$ approach the limit (\ref{cr}),
which has a numerical value about  $ 99$pb, while
$ \sigma_{-\frac{1}{2}-\frac{1}{2} }^{+1\; -1}$ goes with energy as
$ 1/s^3$ and
$ \sigma_{-\frac{1}{2}-\frac{1}{2} }^{+1\;\;\; 0}$ as $ 1/s^2.$

This good high energy behaviour will be violated if the
triple boson coupling differs from its gauge model form.
The cross section starts to increase   with  energy
which leads to a contradiction with  unitarity.
At high energies both  $ \ka$ and $ \la$ terms in
$\sigma_{-\frac{1}{2}-\frac{1}{2} }^{\pm1\;\;\; 0} $ grow with energy
as $ \log s$ and in
$ \sigma_{-\frac{1}{2}-\frac{1}{2} }^{+1\; -1}$ as $ s$ if $ \ka\neq 1$
and $ \la\neq 0.$
The $ \la$ terms in
$ \sigma_{-\frac{1}{2}-\frac{1}{2} }^{-1\; +1}$ increase also with $ s$, while
the cross sections
$ \sigma_{-\frac{1}{2}-\frac{1}{2} }^{\pm1\;\pm1}$ are equal to a constant,
which value  depends only on $ \ka.$
However, these effects would still be quite small at the energies of NLC.
Therefore, it is important to study the sensitivity of  reaction
(1) to the anomalous interaction taking into account the polarization effects.

\section{Anomalous triple boson coupling in the Standard Model}

The anomalous triple boson couplings have been
theoretically well studied in $ e^+e^-$
collisions \cite{hagi}, and   recently also the
beam polarization  has been taken into account in such studies \cite{pan}.
Also the $ e^-e^-$ collisions
have been found to be useful for this purpose \cite{ee}.
One  disadvantage of these collision processes is that
they do not allow separate tests of the
anomalous photon and $ Z^0$ couplings since
both $ \g WW$ and $ ZWW$ vertices are involved in the reactions.
The $ e^-\g$ option to be offered by the NLC will be an ideal tool for
studying the photon anomalous coupling separately.

In the context of
the SM, the  process (1)
has already been investigated previously
 \cite{eg,choi,yehudai,phil}, and its
sensitivity to the photon
anomalous coupling has been found to be comparable
with the estimated sensitivity of
the $ W$ pair production processes \cite{choi}.
In this section we shall  update these  analysis
 for a  $ 500$ GeV $ e^+e^-$ collider,
taking into account the effects of beam  polarization
 and final state polarization as well as
 the recent developments in the linear collider
design.

The scattering of linearly polarized laser light off the electron beam
produces a polarized photon beam with very hard spectrum strongly peaked at
the maximum energy, which is about $ 84$\% of the electron beam energy
\cite{telnov}. The  backscattered photon beam is slightly cone-shaped, where
the hardest photons lie in the center  and the softer components
 form   the outer layers of the cone.
In the $ e^-\g$ collision the precisely collimated
electron beam probes only the
hardest photons of the $ \g$ beam making highly monochromatic collisions
technically feasible\footnote{The author thanks prof. V.Telnov for
a clarifying discussion on this matter.}.
Therefore, we shall carry out our analysis
for the center of mass energy $ \sqrt{s_{ e\g }}=420$ GeV
corresponding to the peak value of the photon spectrum,  assuming that the
small nonmonochromaticity effects of the photon beam
 will be treated  separately in
every particular experiment (similarly to  the treatment of nonmonochromaticity
of the electron beam due to the initial state Bremsstrahlung).
The other relevant  NLC parameters which we have used  are the following:
\begin{itemize}
\item integrated luminosity  $ L_{ int }=100 $fb$^{-1},$
\item the covering region of a detector  $ |\cos\th|\leq 0.95,$
which is already achieved in all LEP experiments,
\item  $ W^-$ reconstruction
efficiency of $ 0.1.$
\end{itemize}

The most straightforward and  experimentally  easiest
observable for testing the parameters $ \ka$ and $ \la$
is the differential cross section.
Since in the SM the
electron beam polarization is fixed by the handedness of the
interaction ($ A=-B=1$), one can vary the initial state by choosing different
photon beam polarization, $ \tau_1=\pm1.$ In order to minimize statistical and
systematic  errors it will be profitable to
sum over the  $ W^-$ polarization states.
Similarly one can study  the total cross section
$ \sigma_{ \tau_1=\pm1 }^{tot}.$
Since  the differential cross sections are strongly peaked in the backward
direction one would expect that also the forward backward asymmetries
\be
A^{ FB }_{\pm}=\frac{\sigma_{ \tau_1=\pm1 }(\cos\theta\geq 0)-
\sigma_{ \tau_1=\pm1 }(\cos\theta\leq 0)}
{\sigma_{ \tau_1=\pm1 }(\cos\theta\geq 0) +
\sigma_{ \tau_1=\pm1 }(\cos\theta\leq 0)}
\label{fb}
\ee
could be sensitive to the anomalous coupling.
The fourth quantity, which reflects the effects of
the beam polarization, is the polarization asymmetry
$ A_{ pol }$ defined as
\be
A_{ pol }(\cos\theta)=\frac{d\sigma_{ \tau_1=+1 }-
d\sigma_{ \tau_1=-1 }}
{d\sigma_{ \tau_1=+1 } +
d\sigma_{ \tau_1=-1 }}.
\label{as}
\ee
We have also studied whether the measurement of the final state
$ W$-boson polarization could offer sensitive tests for $ \ka$ and $ \la.$
The information about the polarization of $ W$-boson can be obtained
by measuring the angular distribution of its decay products.
A suitable quantity would be
the forward-backward asymmetry of the leptons produced in  $ W^-$
decay, which  is related to the cross sections corresponding to the different
$ W^-$ polarization states $ \tau_2=\pm1$
as follows (see \eg ref.\cite{yehudai}):
\be
\chi_{ \pm }^{FB}=\frac{3}{4}\frac{\sigma_{\tau_1=\pm1}^{\tau_2=-1} -
\sigma_{\tau_1=\pm1}^{\tau_2=+1}}{\sigma_{\tau_1=\pm1}^{tot}}.
\label{xfb}
\ee
Let us now study numerically the sensitivity of these five quantities
for determining the values of $ \ka$ and $ \la$ in the NLC.
We have carried out  a $ \chi^2$ analysis
by comparing the SM prediction of the observables with
those corresponding to the anomalous $ \ka$ and $\la.$ All the contours
are calculated at $ 90$\% confidence level, which corresponds to
$ \Delta\chi^2=4.61.$
The statistical errors are computed assuming the NLC parameters
given above. The systematic errors are estimated by assuming
the  uncertainty of the cross section measurement to be
at the level of $ \sim2$\% \cite{sys}, coming mainly from the
errors in the luminosity measurement, the acceptance,
the background subtraction
and the knowledge of branching ratios. For the  asymmetries
the systematic uncertainty corresponding to  the luminosity measurement
cancels and for them the total systematic error of the cross sections
 is taken to be $ 1.5$\%.

All the observables are first   analysed   separately
and after this a combined analysis is performed.
Both forward-backward asymmetries, $ A_{ \tau_1 }^{FB}$ and
$ \chi_{ \tau_1 }^{FB}$, turned out to be several times
less sensitive to the anomalous coupling than the  other three observables
and therefore we shall not present separate results for them.

In Fig. 2 we plot the allowed domains of the photon anomalous
coupling on the $(\ka,\la)$ plane, which are obtained by  analysing the
polarization asymmetry $ A_{ pol }$   (contour $ a$)
and the total cross sections $ \sigma_{ \tau_1=\pm1 }^{tot}$ (contour $ b$).
The combined  contour, resulting from these two, is denoted by $ c.$
As can be seen in  Fig. 2, in this small region
the total cross section is much more
sensitive to the parameter $ \ka$ than to  $ \la.$
This result can be easily
understood on the basis of our discussion in the last Section.
The bulk of the cross section comes
from the polarization states $ \tau_1=\tau_2=\pm1,$ which makes them
 most sensitive to the anomalous coupling. Since
 $ \sigma_{ \tau_1=\pm1}^{ \tau_2=\pm1 } $ do not depend on  $ \la$
at high energies, but depend on $ \ka$ at all energies,
the sensitivity to  $ \la$ is much less than to  $ \ka.$
 If the polarization asymmetry alone would
allow us to measure $ \ka $ with the precision  $ |\Delta\ka|=0.1$
and $ \la$ with the precision about $ |\Delta\la|=0.02,$
 the combined result were much more restrictive, giving
an improvement by a factor of $ 2$ in the $ \la$ measurement
and by a factor of $ 3$ in the $ \ka$ measurement.

The most sensitive observable to the photon anomalous coupling
is the differential cross section.
The contours of  allowed regions in $ (\ka,\la)$ space
 obtained from its analysis  are plotted in Fig. 3.
The curves for the different photon polarization states
$ \tau_1=\pm1$ are indicated in the figure.
The contour resulting from the combined analysis is denoted by $ a.$
As can be seen from Fig. 3 the most stringent constraints for the anomalous
coupling are obtained in the case of  left-handedly polarized
electron and right-handedly polarized photon beams.
This is an expected result, since the s-channel diagram in Fig. 1
does not contribute in this case and the entire  cross section
comes from the u-channel diagram, which probes the triple boson coupling.

The resultant allowed domain for $ \ka$ and  $ \la$
obtained by combining the measurements of  all  five
observables we have considered is denoted by $ b$ in Fig. 3.
As compared with  contour $ a$ the  improvement  achieved by
performing the combined analysis of all observables
 is  small, implying that the main constraints
 come from the measurements of the differential cross sections.

Summarizing, we have found out that by studying reaction (1) in the NLC
with  the assumed set of parameters,
one could  constrain the anomalous triple boson coupling
 parameters $ \ka$ and $ \la$ to the following regions:
\[ -0.008\leq 1-\ka\leq 0.009, \]
\[ -0.01\leq\la\leq 0.005 .\]
Comparison with the recent analysis \cite{pan} of the $ W^-$ pair production in
$ e^-e^+$ collisions with polarized beams
indicates that the sensitivity of a $ \sqrt{s}=500$ GeV
electron-positron collider to  $ \ka$ is similar
to the sensitivity of a
 $ \sqrt{s_{ e\g }}=420$ GeV electron-photon collider (the results are
complementary). The limits
for  parameter $ \la$ would in contrast be more stringent in
$ e^-\g$ than in $ e^-e^+$ collisions.
Let us emphasize again, that the bounds from  process (1) are independent of
the parameters of $ ZWW$ coupling, probing the anomalous coupling of the
photon only.

\section{Single heavy vector boson production in  left-right model}

In this section we shall study a discovery
potential of the final phase of  NLC considering  process (1)
in the framework  of the left-right model.
The gauge boson in the  reaction
$ e^-_R\g\rightarrow W^-_2 N$ is a  heavy right-handed gauge boson
 of the model, while the neutrino could be either a heavy
right-handed neutrino or, in the case of a large neutrino
mixing angle, the ordinary light  neutrino $ \nu$. However, the last
possibility could still be very much suppressed.

In the left-right symmetric model there is a right-handed  triplet Higgs field
$ (\DE^{--},\DE^-,\DE^0),$ which  breaks the
$ SU(2)_R\times SU(2)_L\times U(1)_{ B-L }$ symmetry down to the SM symmetry.
The neutral member of the triplet obtains a vacuum expectation value
$ v$ which is responsible for giving masses to the right-handed
gauge bosons and heavy Majorana neutrinos.
The mass relations are the following: $ M_{W_2}=gv/\sqrt{2}$ and
$ M_N=2hv,$ where $ g$ is the gauge coupling constant of $ SU(2)_R$
(we assume it to be equal to the SM gauge coupling $ g$) and $ h$ is
an unknown Yukawa coupling constant.

The general helicity amplitudes
given in eq. (\ref{hel}) are applicable in this particular
model if we set $ A=B=1,$ \ie, assume purely right-handed interaction.
Then the electron beam should also be  right-handedly polarized.
Parameters $ \ka$ and $ \la$ are assumed to
have their gauge model  value if not stated otherwise.
In order to see the relative importance
 of different polarization states we
present in Fig. 4 the angular distributions of the differential cross sections
for all combinations of polarizations.
In Fig. 4 (I) we plot the differential cross sections for the left-handedly
polarized photon beam assuming the collision energy
$ \sqrt{s_{ e\g} }=1.5$ TeV, the gauge boson mass $ M_{ W_2 }=700$ GeV
and the neutrino mass $ M_N=300$ GeV.
In Fig. 4 (II) we plot the same for the right-handedly polarized
photon beam.
As the figures show,  the final states with a left-handed
neutrino  are clearly suppressed. The main part of the
cross section is again coming from the case where the photon and
$ W^-$ are polarized in the same way. The other polarization combinations
are somewhat suppressed for the right-handedly polarized photon beam.
While in the SM {\em all \/} differential cross sections are peaked in
the backward direction,  in this case the distributions are
more flat for many polarization states providing better possibilities for
detecting the anomalous photon coupling.

  The mass dependence of the  total cross section
of the process $ e^-_R\g\rightarrow W^-_2 N$ can be seen in Fig. 5,
where we plot the cross section as a function of
$ W^-_2$ mass for the center of mass energy
$ \sqrt{s_{ e\g} }=1.5$ TeV assuming the left- (Fig. 5 (I))
and right-handedly (Fig. 5 (II)) polarized photon beams.
The curves denoted by $ a$ and $ b$ correspond to the neutrino masses
 $ M_N=300$ GeV and $ M_N=600$ GeV, respectively.
The cross sections are found to be reasonably large for
almost the entire kinematically allowed
mass region, decreasing faster
with $ M_{ W_2 }$ for the $ \tau_1=1$ photons.
At low $ W_2$ masses the difference between $ a$ and $ b$ curve
is small but for heavy $ W_2$ masses the cross section depends strongly
on the neutrino mass. If $ M_N\leq M_{ W_2 },$  reaction (1) enables us
to study heavier vector bosons than what is possible
in the $ W^-_2$ pair production in
$ e^-e^+$ or $ e^-e^-$ collisions.
The reaction  would be even more useful in this respect
  if the mixing between the heavy and the light neutrino
is large enough to give observable effects.
In Fig. 6 we plot the cross section of the  reaction
$ e^-_R\g\rightarrow W^-_2 \nu$ for different photon polarizations
assuming a vanishing  mass of $ \nu$ and
the neutrino mixing angle of $ \sin\theta_N=0.05.$
For this set of parameters the process should be observable
up to $ W$-boson mass $ M_W=1.2$ TeV.

In order to study how sensitive  the single heavy vector boson production
process (1) is to the photon anomalous coupling in the left-right  model
  we have carried out the
 $\chi^2$ analysis of the differential cross section.
The contours of the allowed regions at $ 90$\% C.L. in $ (\ka,\la)$ space
are plotted in Fig. 7 assuming the collision energy
$ \sqrt{s_{ e\g} }=1.5$TeV, the boson mass $ M_{ W_2 }=700$ GeV and the NLC
parameters given in Section 3.
The resulting curve
for the neutrino mass $ M_N=300$ GeV assuming unpolarized beams
as well as the contours obtained for the
right-handedly polarized electron  and $ \tau_1=\pm1$ polarized
photon beams are indicated in the figure. The combined result
obtained from the measurements with polarized initial states is denoted
by $ a.$ As one can see, the improvement in the sensitivity
compared with the case of unpolarized
beams is essential.

To study how the sensitivity to the anomalous coupling depends on the
neutrino mass we have repeated   the  analysis for the
neutrino mass $ M_N=600$ GeV (contour $ b$ in Fig. 7).
Despite of the smaller cross section the sensitivity
has not decreased significantly. This is a consequence of the $ G_4$ terms
in  helicity amplitudes (\ref{hel}) which are proportional
to the neutrino mass and nonzero only in the presence of anomalous
triple boson coupling.

\section{Summary}

The Next Linear Collider
will provide us with the possibility to study nearly monochromatic high energy
$ e^-\g$ collisions with any combination of polarizations.
This option is particularly suitable for studying the
massive gauge boson production,
since the initial state  photon probes directly
the gauge boson self-interactions vertex $ \g WW.$
We have derived the helicity amplitudes
of the reaction
$ e^-\g\rightarrow W^-N$ assuming arbitrary particle masses and a general
form of the interactions involved.
We have used the general helicity amplitudes to analyse
two  theories: the Standard Model and the left-right symmetric model.

In the case of the Standard Model our numerical results concern
the sensitivity of the first phase of NLC
($ \sqrt{s_{ ee }}=500$ GeV) to the parameters of anomalous photon coupling
$ \ka$ and $ \la,$ taking into account the polarization of the particles.
We have performed a combined $ \chi^2$ analysis of five  observables,
according to which the NLC will be able
to constrain the anomalous coupling parameters
 at $ 90$\% C.L. to the region $ -0.008\leq 1-\ka\leq 0.009, $
$ -0.01\leq\la\leq 0.005.$

The \lr\/ serves as a sample theory for studying the potential of the final
phase of NLC to discover a new heavy right-handed $ W_2.$
The cross section of the reaction is found to be large and,
since there is just one  $ W_2$ in the final state,
the study of this
 process could significantly extend the kinematic region covered by
the $ W$-boson  pair production.
The effects of the photon anomalous coupling are found to be bigger
for the higher neutrino masses, compensating the effects of
smaller cross section in testing  the gauge boson
self-coupling parameters.

\vspace{0.5cm}
\noindent {\bf Acknowledgement.} The author thanks Jukka Maalampi and
Matts Roos for the helpful discussions and
 expresses  his gratitude to
Emil Aaltosen S\"a\"ati\"o, Wihurin Rahasto and Viro S\"a\"ati\"o for grants.

\end{document}